# Progress Toward Superconductor Electronics Fabrication Process With Planarized NbN and NbN/Nb Layers

Sergey K. Tolpygo, *Senior Member, IEEE*, Justin L. Mallek, Vladimir Bolkhovsky, Ravi Rastogi, Evan B. Golden, Terence J. Weir, Leonard M. Johnson, *Senior Member, IEEE*, and Mark A. Gouker, *Senior Member, IEEE*

(*Invited Paper*)

*Abstract*—**In order to increase circuit density of superconductor digital and neuromorphic circuits by 10× and reach integration scale of $10^8$ Josephson junctions (JJs) per chip, we developed a new fabrication process on 200-mm wafers, using self-shunted Nb/Al-AlOx/Nb JJs and kinetic inductors for cell miniaturization. The process has one layer of JJs, one layer of resistors, and ten fully planarized superconducting layers: 8 niobium layers and two layers of high kinetic inductance materials, Mo2N and NbN, with sheet inductance of 8 pH/sq and 3 pH/sq, respectively. The minimum linewidth of NbN kinetic inductors is 250 nm. NbN films were deposited by two methods: with $T_c \approx 15.5$ K by reactive sputtering of a Nb target in Ar+N2 mixture; with $T_c$ in the range from 9 K to 13 K by plasma-enhanced chemical vapor deposition (PECVD) using Tris(diethyl-amido)(tert-butylimido)niobium(V) metalorganic precursor. PECVD of NbN was investigated to obtain conformal deposition and filling narrow trenches and vias with high depth-to-width ratios, $h/w$>1, which was not possible to achieve using sputtering and other physical vapor deposition (PVD) methods at temperatures below 200 ℃ required to prevent degradation of Nb/Al-AlOx/Nb junctions. Nb layers with 200 nm thickness are used in the process layer stack as ground planes to maintain a high level of interlayer shielding and low intralayer mutual coupling, for passive transmission lines with wave impedances matching impedances of JJs, typically ≤50 Ω, and for low-value inductors. NbN and NbN/Nb bilayer are used for cell inductors. Using NbN/Nb bilayers and individual pattering of both layers to form inductors allowed us to minimize parasitic kinetic inductance associated with interlayer vias and connections to JJs as well as to increase critical currents of the vias. Fabrication details and results of electrical characterization of NbN films, wires, and vias, and comparison with Nb properties are given.**

*Index Terms*—**AQFP, Josephson junctions, kinetic inductors, NbN films, niobium nitride, SFQ circuits, superconducting integrated circuit, superconducting integrated circuit fabrication, superconducting thin films, RSFQ**

## I. Introduction

ADVANTAGES of superconductor digital electronics over semiconductor electronics in information processing speed and energy efficiency have been well documented [1]–[2]. The main obstacle to implementation of superconductor electronics (SCE) is its low integration scale and, as a consequence, low functionality in comparison to CMOS-based electronics [3]. For instance, Fig. 1 compares progress in fabrication technology, characterized by the minimum linewidth of the most advanced fabrication process, and integration scale of superconductor and semiconductor electronics as measured by the highest achieved density of active devices, respectively, Josephson junctions and transistors, in demonstrated integrated circuits. The linewidths and transistors count data are from [4] and JJ densities are from [5]–[8].

At present, the junction density, $n_J$ achieved in superconductor circuits is three orders of magnitude lower than the achieved density of transistors, $n_T$ and a 150 nm minimum linewidth achieved in SCE is about 50× larger than in a nominally 3-nm CMOS process. We note that the process node in CMOS industry characterizes the minimum size of the active devices, MOSFETs, whereas in SCE the minimum linewidth pertains to inductors while the minimum size of JJs could be much larger, e.g., 500 nm or 700 nm.

Despite a faster growth of $n_J$ with time in recent years, it is highly unlikely that the SCE circuit density may ever become comparable to the one in CMOS because information encoding by magnetic flux quanta, i.e. loop currents, in SCE requires larger cell areas than charge-based information encoding in CMOS [3]. Nevertheless, we believe that there are many interesting applications for SCE, e.g., digital signal processing, quantum computing, and neuromorphic computing [9], if its integration scale could be increased to about a hundred million JJs per chip, corresponding to over ten million artificial neurons or logic gates if classical computing is concerned.

Therefore, the goals of our fabrication process development and of this work is to increase the circuit density of SCE by at least 10× in comparison with the current state-of the-art of about 1.5×10^7 JJs/cm² [7]. Estimates presented in Sec. II show

Manuscript receipt and acceptance dates will be inserted here. This material is based upon work supported by the Under Secretary of Defense for Research and Engineering under Air Force Contract No. FA8702-15-D-0001. Distribution statement A. Approved for public release. Distribution is unlimited. *(Corresponding author: Sergey K. Tolpygo.)*

All authors are with Lincoln Laboratory, Massachusetts Institute of Technology, Lexington, MA 02421, USA (e-mails: sergey.tolpygo@ll.mit.edu, justin.mallek@ll.mit.edu, blokv@ll.mit.edu, ravi.rastogi@ll.mit.edu, evan.golden@ll.mit.edu, weir@ll.mit.edu, ljohnson@ll.mit.edu, gouker@ll.mit.edu, ).

Color versions of one or more of the figures in this paper are available online at http://ieeexplore.ieee.org.

Digital Object Identifier will be inserted here upon acceptance.

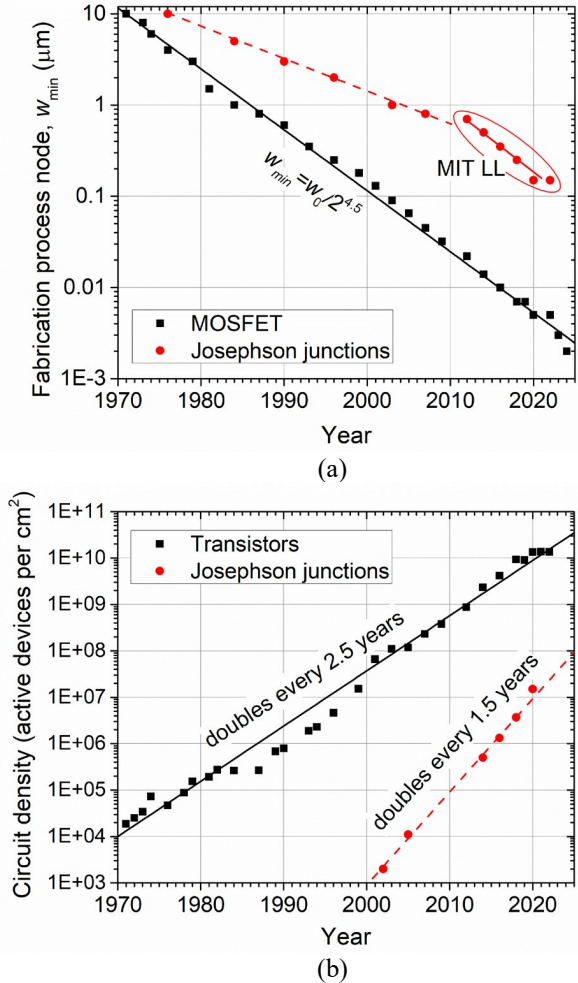

Fig. 1. Progress of superconductor digital electronics in comparison to semiconductor electronics: (a) minimum linewidth of the fabrication process used; (b) number density of the active devices in the demonstrated integrated circuits. In semiconductor industry, the minimum linewidth halved approximately every 4.5 years on average; transistor density doubled approximately every 2.5 years as shown by solid black lines. In recent years, the density of JJs doubles every 1.5 years, but presently is three orders of magnitude below the density of transistors. Note, despite the faster growth, JJ density cannot catch up with transistor density; see text.

that this goal can be achieved by implementing self-shunted Josephson junctions and kinetic inductors in circuit cells in order to significantly reduce their area.

## II. ADVANCED PROCESS LAYER STACK

### A. Balancing Densities of Junctions and Inductors

In SCE, information is encoded by magnetic flux quanta $\Phi_0$ in superconducting loops, i.e. loop currents, and manipulated by switching Josephson junctions in the loops or parametrically changing their flux state. The loops can be characterized by parameter $\beta_L = 2\pi I_c L/\Phi_0$, where $L$ is the loop inductance and $I_c$ is the junction critical current; typically, $\beta_L \sim 2$. On average, one JJ is required per one inductor. In this case, the circuit density is maximized if the number density of inductors, $n_L$ is equal the number density of junctions, $n_J$, $n_L \approx n_J$. Both densities depend on the typical critical current of JJs used in the cells, $\langle I_c \rangle$, and Josephson critical current density, $J_c$, of the junctions used in the fabrication process.

For a single layer of self-shunted junctions, the density is

$$n_J = \chi \left( \left( \frac{4\langle I_c \rangle}{\pi J_c} \right)^{\frac{1}{2}} + \delta \right)^{-2},$$ (1)

where $\chi \leq 1$ is the area filling factor, $\delta$ is the minimum surround of the junction top electrode by the bottom electrode in Nb/Al-AlO$_x$/Nb Josephson junctions; it is a process-dependent parameter, currently $\delta \geq 150$ nm.

At $J_c \geq 600$ μA/μm$^2$, Nb/Al-AlO$_x$/Nb junctions become sufficiently self-shunted to use without external shunt resistors. Therefore, $J_c =600$ μA/μm$^2$ is the target current density in our advanced process node. The typical current $\langle I_c \rangle$ depends on the circuit, acceptable bit error rate, energy dissipation requirements, and other factors. For circuits operating at 4 K, it can be as low as 10 μA, e.g., in neuromorphic circuits processing information stochastically [10]–[12], about 25 μA in RQL circuits [13], about 50 μA in Quantum Flux Parametron (QFP) circuits [14], and as high as $\sim$ 175 μA in RSFQ [15] and ERSFQ [15] circuits if extremely low bit error rates are required. The plot of (1) is shown in Fig. 2.

Area of the typical inductor depends on the linewidth, $w$ and the minimum spacing, $s$ between the inductors, the size and number of vias connecting the inductor, and on the type of inductors and their inductance per unit length, $L_l$. Typically, one side of an inductor is connected to a JJ and another side to the ground plane or something else requiring, on average, one via with area $A_{via}$. The typical inductor number density on one superconducting layer is

$$n_L = \chi \left( \frac{\beta_L \Phi_0}{2\pi \langle I_c \rangle L_l} (w + s) + A_{via} \right)^{-1}.$$ (2)

In the existing fabrication processes, $s \geq 250$ nm and the minimal $A_{via} \approx 0.36$ μm$^2$. The minimum spacing between superconducting wires is determined by the process capabilities to fill in the gap between them by SiO$_2$ at temperatures below $\sim$ 200 ℃, and also by mutual inductance between closely spaced inductors [16]. It is not expected to change in the new process nodes.

### TABLE I
### INDUCTANCE PER UNIT LENGTH OF STRIPLINE INDUCTORS

| Inductor / Linear inductance | $L_l$ (pH/μm) $w$ =150 nm | $L_l$ (pH/μm) $w$ =250 nm | $L_l$ (pH/μm) $w$ =350 nm |
|---|---|---|---|
| 200-nm Nb, M6aM4bM7[a] | 0.7475 | 0.5674 | 0.4716 |
| 200-nm Nb, M6aM4bM8[b] | 0.8376 | 0.6563 | 0.5593 |
| 200-nm NbN, M6aM4bM7[c] | 10.507 | 6.423 | 4.654 |
| 100-nm NbN, M6aM4bM7[c] | 20.605 | 12.482 | 8.982 |
| 200-nm NbN, M6aM4bM8[c] | 10.597 | 6.512 | 4.742 |
| 100-nm NbN, M6aM4bM8[c] | 20.695 | 12.571 | 9.070 |

[a] SFQ5ee process: all layers are Nb with thickness 200 nm, dielectric thickness between M6 and M4 is $d_1$ = 615 nm, and between the ground planes $H$ = 1015 nm.

[b] Advanced process SFQ7ee: all layers are Nb with thickness 200 nm, dielectric thickness between the M6 and M4 is $d_1$ =1015 nm, and between the ground planes M4 and M8 is $H$ = 1415 nm.

[c] In the SFQ7ee node, Nb layer M6 can be replaced by a 200-nm NbN layer or NbN/Nb bilayer with 100 nm /100 nm layer thicknesses; see text.

The typical values of $L_l$ for Nb stripline inductors M6aM4bM7, standing for signal traces on layer M6 above Nb ground plane M4 and below Nb ground ("sky") plane M7, are given in Table I based on the data in [16]–[18] for the current fabrication process node SFQ5ee with 8 niobium planarized layers on 200-mm wafers, described in [19], [20]; see [19, Fig. 1], [10, Fig. 1].

In the advanced fabrication process, which we named SFQ7ee, we add an additional Nb superconducting layer M8 above the layer M7 in the SFQ5ee process. This frees up the layer M7 for defining inductor traces, giving potentially two layers of inductors (M6 and M7) above the layer of JJs, whereas M8 becomes the upper ground (sky) plane. Linear inductance of M6aM4bM8 striplines for this process is also given in Table I, calculated using formulas given in [16].

The density of inductors following from (2) and Table I is shown in Fig. 2 by dash-dot curves for Nb striplines inductors M6aM4bM7 with linewidth, respectively from bottom to top, 350 nm, 250 nm, and 150 nm. Shown by the dash curve is the density of Nb stripline inductors M6aM4bM8 with $w$=150 nm. It can be seen that, regardless of the linewidth of Nb inductors, the circuit density is completely determined by the area occupied by inductors in the entire range of critical current of interest. This conclusion agrees with the results of the actual circuit designs with self-shunted JJs and fabrication [7], [8], showing that JJs occupy less than 11% of the circuit area, whereas most of the circuit area is taken by inductors and flux transformers; see Fig. 3 as an example of the densest SCE circuit fabricated thus far.

Hence, using Nb inductors, we cannot reach densities of $10^8$ devices per cm²; the circuits will always be "starved" by inductors. Instead, we need to use a material with much larger linear inductance, e.g., materials with penetration depth much larger than the film thickness, $\lambda \gg t$, providing a large kinetic inductance, $L_{Kl} = \mu_0\lambda^2/(tw) = 1.257\lambda^2/(tw)$ in pH/µm for $\lambda, t$ in µm, as was proposed in [3], [21], [22].

From the fabrication stand point, it is convenient to use nitrides of transition metals, e.g., NbN, NbTiN, Mo₂N which could be deposited both by reactive sputtering and by chemical vapor deposition. Deposition of NbN films will be described in Sec. III. The linear inductance of NbN striplines M6aM4bM7 with Nb ground planes M4 and M7 was measured in [27], and the penetration depth in reactively sputtered films was found to be $\lambda_{NbN}$=491 nm. Using these data, given also in Table I, and (2), the density of NbN stripline inductors is shown in Fig. 2 by four solid lines for NbN signal trace width $w$= 250 nm and three thicknesses: 200 nm, 100 nm, and 50 nm, respectively from bottom to top. The fourth, uppermost, solid line is for NbN striplines with $w$=150 nm, $t$=50 nm, and a significantly smaller via area.

We see from Fig. 2 that balancing densities of inductors and JJs at about $10^8$ devices per cm² level is possible using a single layer of NbN inductors with the film thickness $t \le 100$ nm and linewidth $w \le 250$ nm, and JJs with critical currents in the range from ~100 µA to 150 µA. However, achieving harmony at lower $\langle I_c \rangle$ values or reaching densities significantly above $10^8$ cm⁻² requires either multiple layers of inductors or a significant reduction of the linewidth, below 150 nm, and of the via area. The former is problematic because of the insufficient current carrying capacity of ultranarrow lines of kinetic inductors, which is inversely proportional to their kinetic inductance [21]. The latter requires filling high aspect ratio (depth to width) vias, that is not possible using physical vapor deposition (PVD) methods at deposition temperatures below 200 °C, required for the thermal stability of AlOx barrier in JJs.

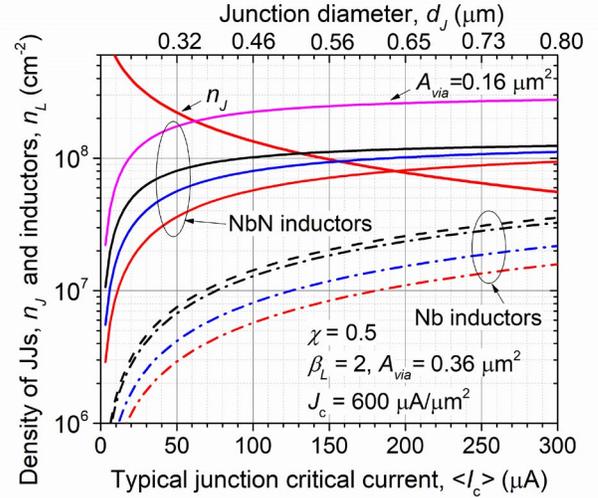

Fig. 2. Dependences of the densities of Josephson junctions, $n_J$ (1) and inductors, $n_L$ (2) on the typical critical current ($I_c$) of JJs in the information processing cells, assuming: $\beta_L = 2$, $\delta = 150$ nm, $s = 250$ nm, area filling factor $\chi = 0.5$, via area $A_{via}$=0.36 µm² in (2), and self-shunted JJs with $J_c = 600$ µA/µm²; diameter of the JJ is given by the top axis scale. Three bottom dash-dot curves are for Nb stripline inductors M6aM4bM7 with linewidth, respectively, 350 nm, 250 nm, and 150 nm from bottom to top, and dielectric thicknesses corresponding to the SFQ5ee process. The fourth, dash curve is for Nb striplines M6aM4bM7 with $w$=150 in the SFQ7ee process. Solid curves correspond to NbN stripline inductors with $w$=250 nm and thickness, from bottom to top, 200 nm, 100 nm, and 50 nm. The uppermost solid magenta curve corresponds to NbN striplines with $w$=150 nm, $t$=50 nm, and $A_{via}$=0.16 µm² in (2).

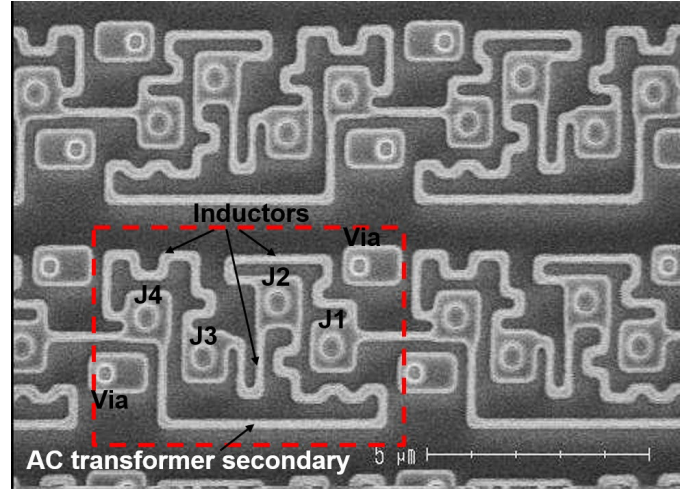

Fig. 3. Scanning electron microscope (SEM) picture of a part of an ac-biased shift register after fabrication of the self-shunted junctions with $J_c$ =600 µA/µm² and stripline-type inductors interconnecting JJs on niobium layer M6. The shift register unit cell is shown in the red dash rectangle. It is repeated thousands of times to form the shift register. All white-grey lines with bends are Nb inductors. Inductor bends are needed to accumulate the required length and provide the proper inductance values. Bottom layers containing the bottom electrodes of the JJs and the transformer primary are not visible. Shift register circuit diagram and operation are described in [24], [6]. Fabrication process used was described in [7]. Inductor linewidth is 150 nm. Photolithography of the inductors was done using an ASML scanner with 193-nm exposure wavelength. The unit cell dimensions are 7 µm × 4 µm. The circuit density is 1.3×10⁷ JJ/cm²; it corresponds to the last datapoint in Fig.1b. In this circuit, the junctions occupy about 11% of the cell area, interlayer vias occupy about 13%, flux trapping moats in M4 and M7 ground planes occupy about 6% of the area. The remaining ~70% of the total area is occupied by the cell inductors and the ac power transformer. Moving the inductors closer to each other to reduce the amount of "dark space" increases parasitic coupling between the inductors.

Since inductance of NbN striplines M6aM4bM8 and M7aM4bM8 is nearly the same and dominated by kinetic inductance of the strips, we can utilize both layers to balance circuit density of circuits with $\langle I_c \rangle \sim 50$ μA. However, this also requires reliable fabrication of JJs with diameters $d_J \approx 0.32$ μm at $J_c = 600$ μA/μm², which is at the border or beyond capabilities of the existing processes [23], [30]. Therefore, we suggest to set the next process node at $d_J$ in the range from 0.5 μm to 0.6 μm, corresponding to the intersections of the black and blue solid curves of the $n_L(I_c)$ dependences (2) in Fig. 2 with the $n_J(I_c)$ dependence (1).

We note that there are circuits, e.g., using Adiabatic Quantum Flux Parametrons [2], which do not require shunted junctions. Critical currents of about 50 μA, convenient for their design and operation, can be obtained using the standard $J_c = 100$ μA/μm² in the SFQ5ee process and $d_J \approx 0.8$ μm. In this case, however, the achievable junction density is at least six times lower and theoretically can be balanced by the density of NbN inductors at about $6 \times 10^7$ cm⁻² level; see also [22].

### B. SFQ7ee Process Layer Stack and NbN/Nb Bilayers

Implementing kinetic inductors can significantly increase SCE circuit density. However, all Nb layers in the process stack cannot be replaced by NbN or other kinetic inductors having much larger penetration depth than $\lambda_{Nb} = 90$ nm because this would dramatically compromise screening properties of the ground planes, increase parasitic coupling between inductors, increase impedance of and propagation delays in passive transmission lines (PTLs), etc. Therefore, materials with high kinetic inductance can be used only on some dedicated inductor layers closest to the JJ layer, e.g., M6 and M7, while Nb layers must be preserved for ground planes, and data and clock routing. Based on this consideration, we propose a layer stack shown schematically in Fig. 4 for the advanced process SFQ7ee.

In comparison to the current process SFQ5ee, we added an additional Nb superconducting layer, M8, and introduced NbN kinetic inductors on the layer M6, which is deposited as NbN/Nb bilayer. Depending on the circuit design requirements, layer M7 in the stack can be either Nb layer, as in the SFQ5ee process, or a NbN/Nb bilayer similar to the layer M6.

The reasons and motivation for using NbN/Nb bilayers, i.e., using a layer of geometrical inductors on top of a layer of kinetic inductors, was given in [22], [27]. Simply, many types of superconductor logic and memory, especially those using ac excitation, rely on transformers in their operation. Replacing niobium geometrical inductors by kinetic inductors reduces length of the inductors, which decreases their mutual running length, and mutual inductance, and reduces efficiency of the transformers. The use of NbN/Nb bilayers allows us to combine kinetic and magnetic (geometrical) inductors on the same circuit level and thereby preserve the required mutual inductance in transformers and provide compact kinetic inductors in logic cells [22]. This is achieved by pattering the bilayer layers individually using two separate photolithography and etching steps.

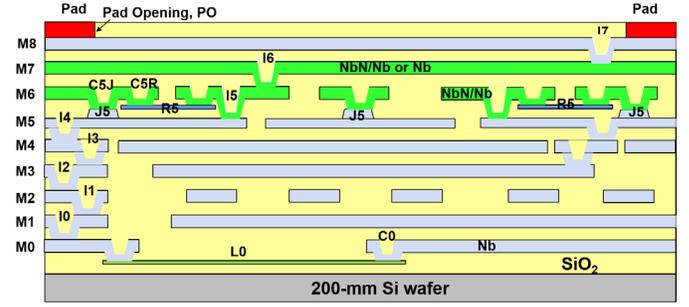

Fig. 4. Cross section of the advanced fabrication process node SFQ7ee. The layer stack from the bottom layer of bias kinetic inductors, L0, up to the resistor layer, R5, is identical to the MIT LL process SFQ5ee [19], [20]. In the new process node, layer M6 is deposited as a bilayer NbN/Nb. Layer M7 can be either NbN/Nb bilayer or a pure Nb layer, depending on the circuit design requirements. An extra Nb layer M8 was added in comparison to the SFQ5ee. The total number of superconducting layers is 10; one layer of resistors, one layer of Nb/Al-AlOx/Nb Josephson junctions, J5. The junctions and resistors are contacted by the layer M6 through vias C5J and C5R, respectively. Vias between Nb layers, I0, I1, etc. are etched in the planarized SiO2 interlayer dielectric with the same name, I0, I1, etc. and are filled by the superconducting metal of the next layer.

The idea of using multilayers as circuit layers in superconductor integrated circuits in order to tailor properties of the individual layers was suggested in [25]. Proximity-coupled S/N bilayer and multilayers are frequently used to tailor the critical temperature of the resultant multilayer, e.g., in transition edge sensors; see e.g., [26] and references therein. Note, that our purpose and, as a result, fabrication process is completely different because we want to preserve the unique properties of the individual layers in the bilayer.

## III. Fabrication Process With NbN Layers

### A. NbN Deposition by Reactive Sputtering

NbN films were deposited in a sputtering chamber of an Endura PVD cluster tool using reactive sputtering of a 13-inch diameter Nb target in N2+Ar mixture at deposition temperatures from room to 400 °C. The deposition power, total pressure and the $N_2$/Ar gas flow ratio were adjusted to optimize the superconducting critical temperature, $T_c$, film thickness and sheet resistance uniformity across 200-mm wafers, and produce compressive residual stress in the films. The typical ranges of the deposition parameters used in the process optimization are given in Table II. Three film thicknesses, 50 nm, 100 nm, and 200 nm, were targeted for applications in NbN/Nb bilayers and as a single layer.

TABLE II
THE TYPICAL REACTIVE SPUTTERING DEPOSITION PARAMETERS OF NbN FILMS

| Parameter (Units) | Value / Range |
|---|---|
| Deposition power (kW) | 1.0 – 2.0 |
| Total deposition pressure range (Pa) | 0.44 – 0.48 |
| N2/Ar flow ratio (sccm/sccm) | 20/70– 30/60 |
| Deposition rate (nm/s) | 0.5 –0.9 |
| Film thickness (nm) | 50, 100, 200 |
| Sheet resistance at 300 K of 200-nm films (Ω) | 7 – 20 |
| Wafer-averaged residual stress (MPa) | −300 |
| Critical temperature (K) | 14.2 – 16.2 |
| Magnetic field penetration depth in 200-nm films (nm) | 491±5 [27] |

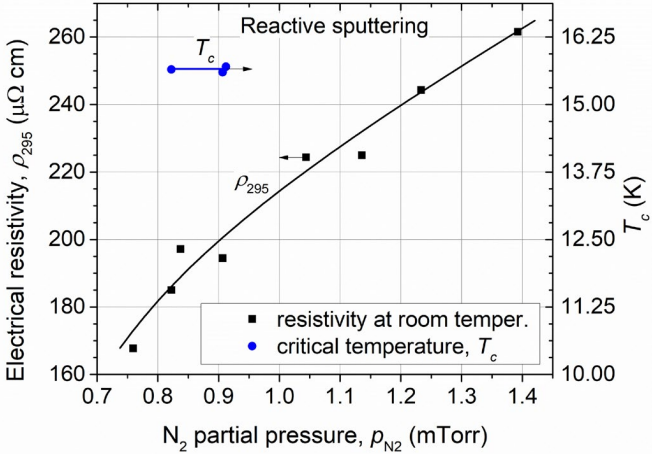

Fig. 5. Resistivity of NbN films with thickness of about 200-nm at room temperature, $\rho_{295}$, as a function of nitrogen partial pressure, $p_{N2}$, during reactive sputtering at a fixed magnetron sputtering power of 1500 W and deposition temperature $T_d$ =200 °C. (left axis). Here $p_{N2} = p_{tot} N_2/(N_2 + Ar)$, $p_{tot}$ is the total deposition pressure, and $N_2/(N_2 + Ar)$ is the ratio of the nitrogen flow to the total flow of nitrogen and argon during the deposition. Also shown in the superconducting critical temperature, $T_c$, of the obtained films (right axis) in the range of partial pressures used for depositing NbN layers in the multilayer process described in this work.

Process flow of the SFQ5ee process with 8 niobium planarized layers on 200-mm wafers, described in [19], [20], was followed up to the metal layer M6, the layer interconnecting Josephson junctions and shunt resistors; see [19, Fig. 1], and [20, Fig. 1]. At this step, a NbN layer was deposited instead of Nb. After the standard 248-nm photolithography, NbN layer M6 was etched using high-density plasma etching in $Cl_2$-based chemistry. The patterned layer was planarized using $SiO_2$ deposition and chemical mechanical polishing, forming interlayer dielectric, I6. Then, Nb layer M7 was deposited and patterned as in the standard SFQ5ee process, followed by the wafer surface passivation and contact pad metallization steps completing the process flow. Obtained circuits and process control monitors (PCMs) were tested at room temperature, using a semi-automated wafer prober, and in liquid He after dicing the fabricated wafers into individual chips.

Dependence of electrical resistivity of the NbN films at room temperature, $\rho_{295}$, and superconducting critical temperature, $T_c$, on the nitrogen partial pressure, $p_{N2}$, during reactive sputtering at a fixed dc magnetron power and 200 °C deposition temperature is given in Fig. 5. Based on the $T_c$, thickness uniformity, and stress optimization, the partial pressure range around 0.85 mTorr was selected for the process. NbN films produced at this pressure have a small variation in the composition across the wafer from 52.5 at. % Nb and 47.5 at. % N in the center of the wafer to 53.5 at. % Nb and 46.5 at. % N at the periphery of the wafer, as measured by the energy dispersive x-ray spectroscopy (EDS). In all cases, NbN films with the highest $T_c$ values were nitrogen deficient, corresponding to $NbN_{1-x}$ with $x \approx 0.1$.

### B. PECVD of NbN From Metal-Organic Precursor

One of the drawbacks of PVD methods is inability to fill narrow trenches and vias with high ratio of depth, $d$ to width $w$, $d/w > 1$, etched in interlayer dielectric as will be explained in more detail in III.C. Therefore, we developed plasma-enhanced chemical vapor deposition (PECVD) of NbN because we are not aware of any CVD methods to deposit Nb which would be compatible with the Josephson junction process.

High-temperature CVD and PECVD processes using $WF_6$ to create tungsten plugs (vias) between metal wiring layers had been successfully used in semiconductors industry for many years until they were replaced by electroplating of Cu. $NbF_5$ is a volatile compound and theoretically could be used to deposit Nb using a CVD process. However, formation of HF as a result of $NbF_5$ reduction by $H_2$ and high temperatures required for the process are damaging for Nb and $SiO_2$ layers existing on the wafers. Instead, we relied on the prior work [28], [29] showing that superconducting transition metal nitrides and carbo-nitrides can be deposited at relatively low deposition temperatures, using metalorganic precursors.

A 200-mm-wafer-capable PECVD system was custom-specified and manufactured by PlasmaTherm, Inc. It utilizes volatile metal-organic precursors to produce highly conformal thin films. Two on-board precursor delivery systems, one delivering Tris(diethylamido)(tert-butylimido)niobium(V), TBTDEN, and another one delivering Tetrakis(diethylamino)titanium(IV), TDEAT, are used for deposition of NbN, TiN, and NbTiN. Hydrogen, nitrogen, and ammonia reactant gasses enable thermal and plasma-based deposition of high-quality films. The system uses a dual high-frequency (HF) and low-frequency (LF) plasma source, fast cycling valves, and fast matching high frequency network to enable ultrashort deposition cycles.

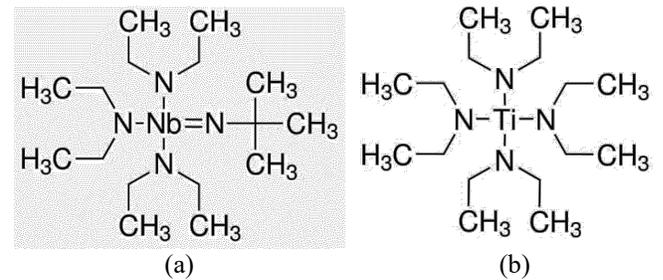

Fig. 6. Metal-organic precursors used to deposit NbN, TiN, and NbTiN films using PECVD: (a) Tris(diethylamido)(tert-butylimido)niobium(V), TBTDEN; (b) Tetrakis(diethylamino)titanium(IV), TDEAT. PECVD process uses 3-to-1 mixture of $H_2$ and $N_2$, a dual frequency plasma source (HF and LF bias), and deposition temperatures from 150 °C to 400 °C. Two on-board precursors in the PECVD system used allow for depositing NbN, TiN, and mixed NbTiN films.

NbN depositions were done using dual-frequency plasma consisting of TBTDEN and 3:1 ratio of $H_2$ to $N_2$ at temperatures from 150 ºC to 400 ºC. Chemical composition of the typical PECVD films is given in Fig. 7.

Dependence of superconducting critical temperature, $T_c$ of NbN films on the deposition temperature is shown in Fig. 8. $T_c$ increases up to 13 K with increasing the deposition temperature to 375 ºC, but in all cases remains lower than $T_c$ of the reactively-sputtered NbN films of about 16 K. This may be caused by some degree of contamination of PECVD films by carbon; see Fig. 7. The films deposited at relatively high temperatures $T_d \gtrsim 300$ ºC have a metallic dependence of electrical resistance, $R$, on temperature with $RRR = R(295 \text{ K})/R(20 \text{ K}) \geq 1$; the films deposited at lower temperatures demonstrate $RRR < 1$, apparently due to a higher degree of disorder and effects of electron localization. Further optimization of the deposition process may be required. Nevertheless, starting from the deposition temperature of 175 ºC, the $T_c$ of the obtained NbN films is comparable or higher than the $T_c$ of Nb. Hence, NbN PECVD films deposited at or slightly below 200 ºC can be used in our multilayered process as kinetic inductors.

## C. Filling Narrow Trenches and High-Aspect-Ratio Vias: Damascene Processing

As was explained in Sec. II, another factor limiting SCE circuit density is area occupied by interlayer vias. Currently, the minimum size of the vias etched in the interlayer dielectric with $d = 200$ nm thickness is about 350 nm, i.e., the aspect ratio 1:1.75 or 0.71:1, and limited by the ability to fill them with superconducting metal of the next metal layer. With decreasing the via size $w$, the depth to width ratio, aspect ratio $d/w$, increases causing shadowing of the bottom of the via during the metal deposition. That is, different points on the via surface experience different deposition rates depending on their visibility from the metal deposition source. As a result, the metal grows faster at the top of the via, near the via entrance, and much slower on the via walls near the bottom. Eventually, the top of the via becomes "clogged," leaving a void inside the vias and forming a so-called key-hole structure; see Fig. 9.

The described problem is exactly the same as the problem of filling narrow gaps between etched metal lines, i.e., inductors, by the interlayer dielectric. At low deposition temperatures, the sticking coefficient of the incoming $SiO_2$ flux is high and their surface mobility is low; therefore, the film growth profile at each point is determined by the point visibility. We use a high-density plasma and apply high RF bias power to create bombardment of the growing $SiO_2$ film by Ar ions, causing re-sputtering of the growing film near the top of the metal gaps, densification of the film, and improving the gap fill. Nevertheless, at the gap depth to width ratios above about 200 nm to 250 nm, i.e. 0.8:1, a void in the dielectric starts to form. This limits the minimum spacing $s$ between metal lines in the process to 250 nm. The limiting aspect ratio for metal gaps is higher than for the vias because gaps (trenches) being two-dimensional objects have visibility restricted only mainly from two sides while vias are shaded from all four sides.

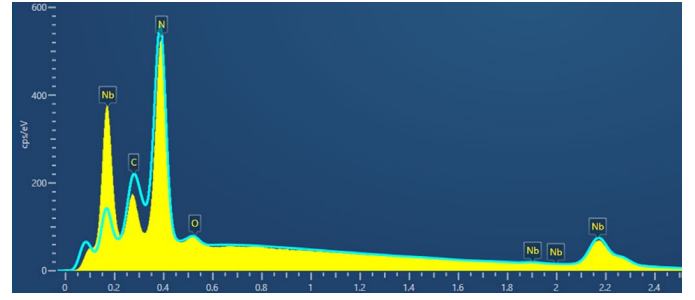

Fig. 7. Energy dispersive x-ray spectroscopy (EDX) of PECVD NbN. The blue curve is the best fit to elemental composition of PECVD films: 44% niobium, 43% nitrogen, ~12% carbon, and 1% oxygen. Note that carbon fitting is poor, overestimating the amount of carbon in the film.

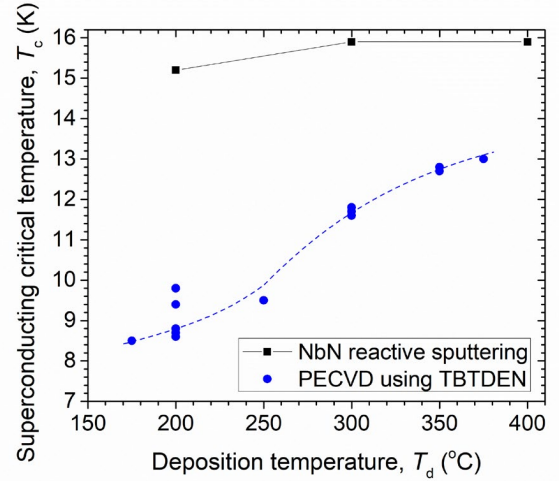

(a)

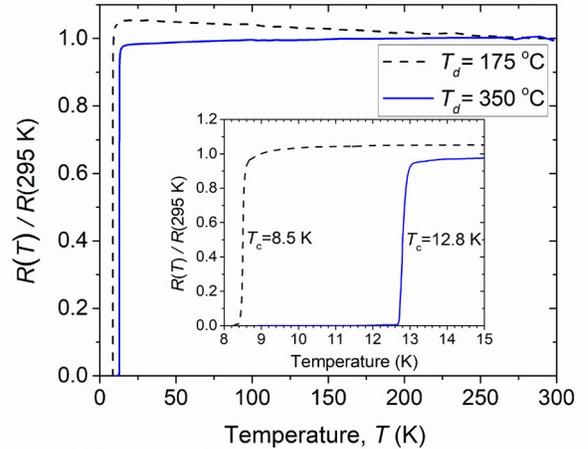

(b)

Fig. 8. (a) Critical temperature $T_c$ of NbN films as a function of deposition temperature, $T_d$ : (●) PECVD films were produced using TBTDEN metal-organic precursor and $H_2+N_2$ plasma at a fixed HF source power of 750 W; (■) reactively sputtered NbN films at dc magnetron power of 1500 W and $p_{N2}$ = 0.838 mTorr (0.112 Pa); point connections are to guide the eye. (b) Resistance $R$ vs temperature $T$ dependence of NbN PECVD films deposited at 175 ºC and 350 ºC. Inset: zoom in the superconducting transition region.

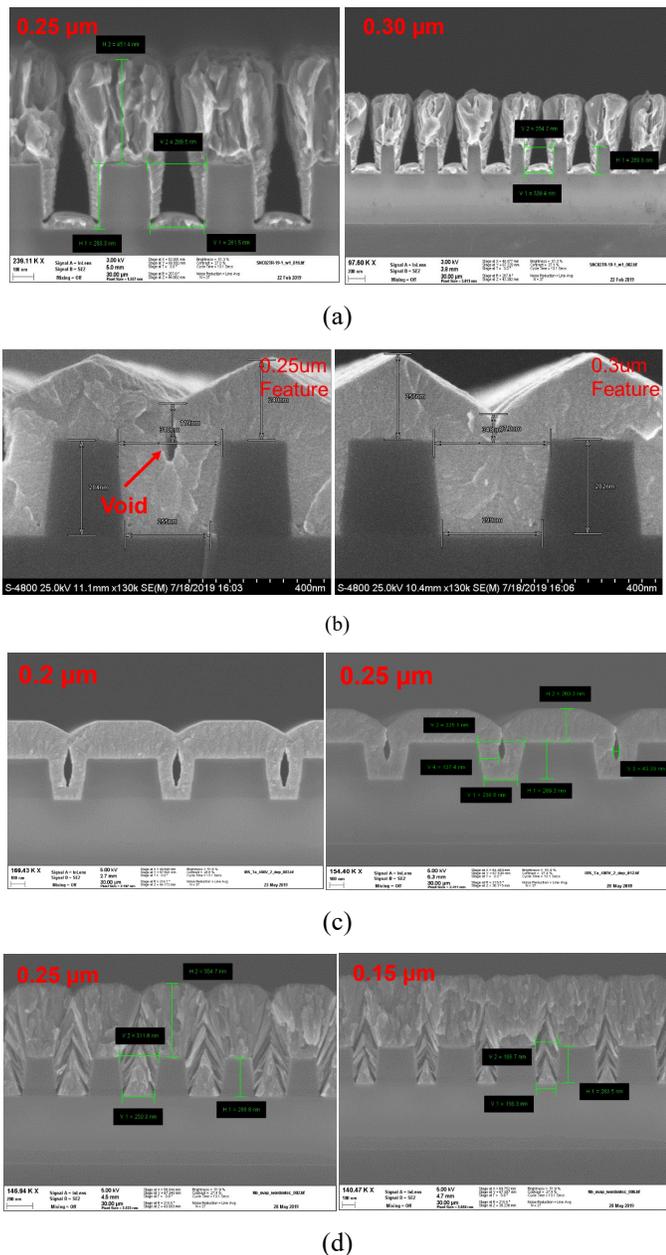

Fig. 9. SEM images of the cross sections showing results of the attempts to fill narrow trenches etched in the SiO$_2$ interlayer dielectric by Nb film, using various PVD methods: (a) magnetron sputtering in an Endura deposition system; (b) Ionized metal plasma deposition SIP EnCoRe process (Applied Materials, Inc.); (c) argon ion-assisted deposition; (d) e-beam evaporation. All trenches were about 240 nm deep; the width of the trenches is shown in the SEM pictures. In all images, Nb metal is grey or light grey, SiO$_2$ is dark grey. Voids forming in the trenches and/or Nb films are clearly seen in all images at trench widths below about 300 nm, aspect ratios above 0.8:1, depending on the deposition method.

In semiconductor electronics manufacturing, the problem of metal gap filling is solved using either much higher deposition temperatures than we can use without degrading AlO$_x$-based Josephson tunnel junctions, providing much higher mobility of the depositing species, or by changing the entire process to the so-called damascene processing schematically shown in Fig. 10.

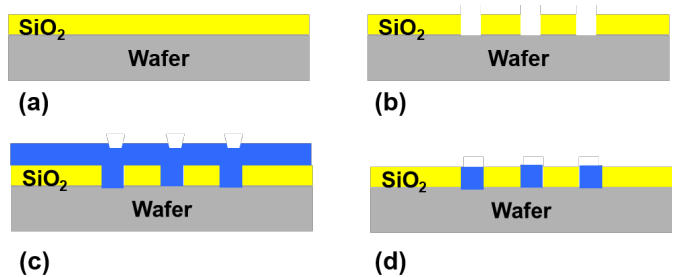

Fig. 10. Metal damascene process for forming narrow trenches and vias with high aspect ratio: (a) interlayer dielectric deposition on the planar surface; (b) etching narrow trenches where metal lines need to be formed, or vias; c) filling the etched trenches or vias with (PE)CVD metal; d) metal planarization by chemical-mechanical polishing to form a planar surface for the next processing step.

The metal damascene process to form narrow metal lines and vias is an inverted process to the one used in SCE. In the current process a deposited metal layer is etched, an interlayer dielectric is deposited filling gaps, then the layer topography is removed by using the dielectric chemical-mechanical polishing (CMP). Then vias are etched and filled by the next metal layer. In the metal damascene process, the dielectric is etched first, Fig. 10b, then a metal layer is conformally deposited to completely fill in the narrow trenches or vias etched in the dielectric, Fig. 10c. Then, the metal on the flat surfaces is etched away by the metal CMP to isolate the metal lines as shown in Fig. 10(d). The next processing step follows. In the simplest case of vias, this is the next metal layer deposition. In a more general case, this could be the next dielectric layer deposition to repeat the damascene processing of the next metal layer. The enabling technology for the damascene processing is conformal deposition of the metal layer over narrow trenches, which is usually accomplished using metal CVD at high temperatures or electroplating. Unfortunately, neither process exists for superconducting materials.

We investigated all PVD methods available to us to fill in narrow trenches by Nb: magnetron sputtering, biased magnetron sputtering forming ion-metal-plasma, electron-beam (e-beam) evaporation, and argon-ion-assisted deposition. Cross sections of some of the deposited films are shown in Fig. 10. It is easy to see that all the used PVD methods failed to fill in trenches narrower than about 250 nm without forming voids in the direction along the trenches; in the case of e-beam evaporation, the voids also formed in the film due to preferential growth of Nb grains at 45 degrees to the SiO$_2$ surfaces of the trenches.

Therefore, we have developed conformal deposition of NbN films over narrow trenches and into small vias using PECVD. The minimalist goal is to produce NbN filled vias with sizes below the current Nb via size of 350 nm. The maximalist goals are to develop the whole damascene process using NbN conformal deposition and CMP as in Fig. 10.

The results of this development are shown by cross sections in Fig. 11. Using a sloped sidewall profile, we were able to fill trenches with aspect ratio as high as 2.85:1 on trenches with width changing from 70 nm to 170 nm; see the top panel in Fig. 12a. We have also demonstrated filling vias with sizes down to 140 nm; see Fig. 12b. Using the vertical sidewall profile, we

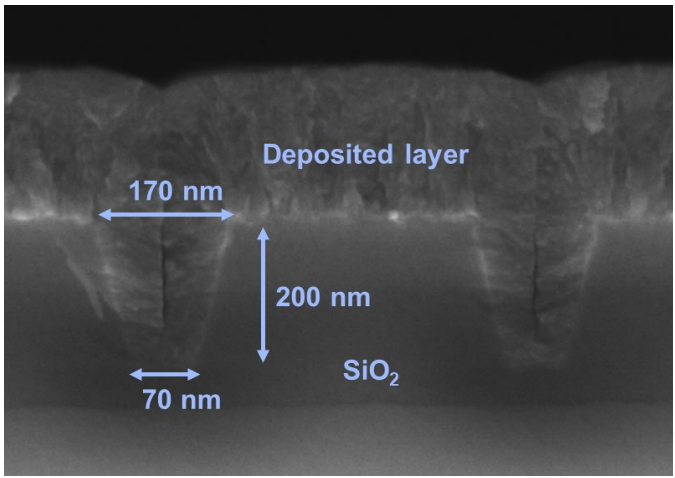

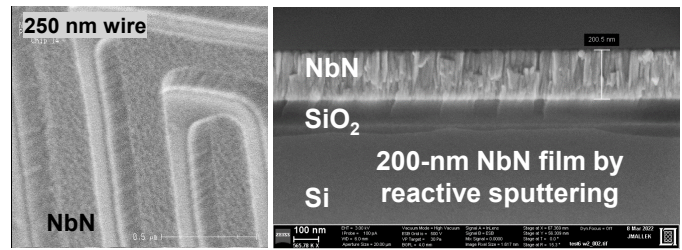

Fig. 13. SEM tilted to view of a reactively sputtered NbN patterned into snakes (meanders) using 248-nm photolithography and high-density plasma etching (left image); the film cross section (right image).

Further improvements in the topography and reduction of the width of vias and trenches requires further optimization of the PECVD process and is on-going.

### D. NbN Etching and Chemical-Mechanical Planarization

NbN etching was done using high-density plasma (HDP) and BCl$_3$/Cl$_2$/Ar (20/90/10 sccm) gas mixture at 10.7 Pa pressure; Inductively coupled plasma (ICP) source power of 700 W, and 13.6 MHz RF bias power of 70 W. The typical 250 nm lines and spaces are shown in Fig. 13.

NbN CMP process was developed using silica-based slurry with H$_2$O$_2$ (8:1 dilution) at pH ~ 2.5, which is a modified industry-standard recipe for tungsten CMP.

### E. NbN/Nb Bilayer Deposition and Etching

Similar to II.A, the layer M6 was deposited in-situ as NbN/Nb bilayer with total thickness of 200 nm or 250 nm. Thicknesses of the individual layers $t_{NbN}$ and $t_{Nb}$ were varied to investigate various combinations, e.g., NbN(150)/Nb(100) and NbN(100)/Nb(100).

Patterning of the bilayer was discussed in [22], [27]. It was done in two photolithography and etching steps to pattern individually the top Nb and the bottom NbN layers of the bilayer. The first photolithography step uses a dark field mask M6a and positive photoresist (PR). This step does not exist in the standard SFQ5ee process, and creates unmasked etch windows for were able to fill trenches with width down to 170 nm; see the bottom panel in Fig. 12a. Due to insignificant topography obtained, the films deposited over the vias can be patterned even without the CMP step, which simplifies the fabrication process.

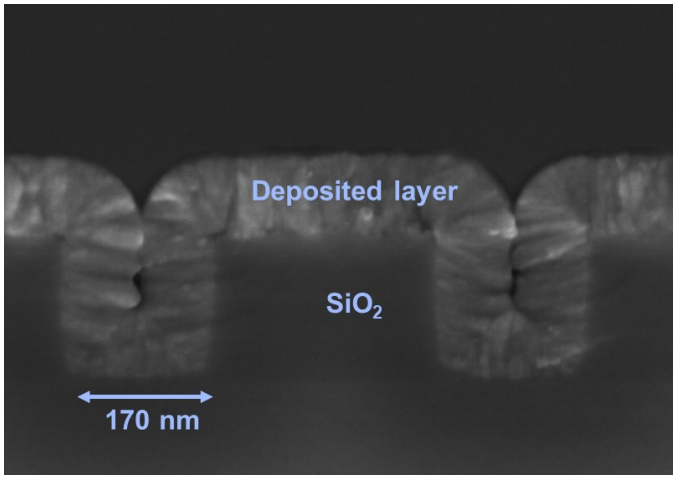

(a)

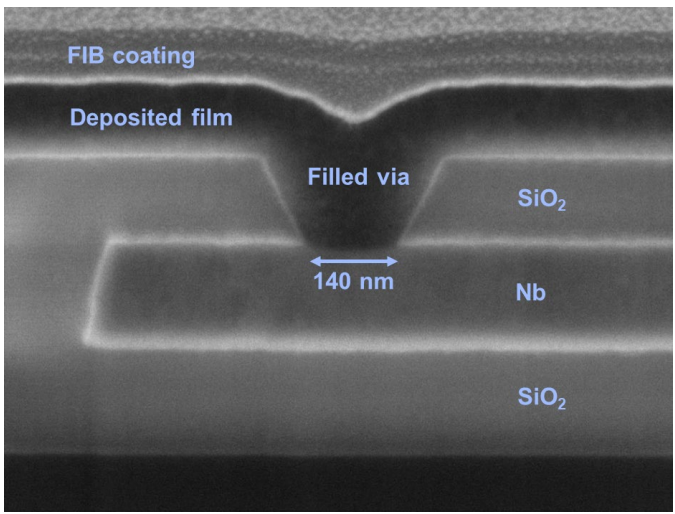

(b)

Fig. 12. SEM images of cross sections showing: (a) conformal PECVD of NbN over trenches in SiO$_2$ with aspect ratio 2.85:1 using sidewall sloping (the top image), and with aspect ratio 1.2:1 and vertical sidewall (the middle image). (b) Conformally filled 140-nm via with sloped sidewalls.

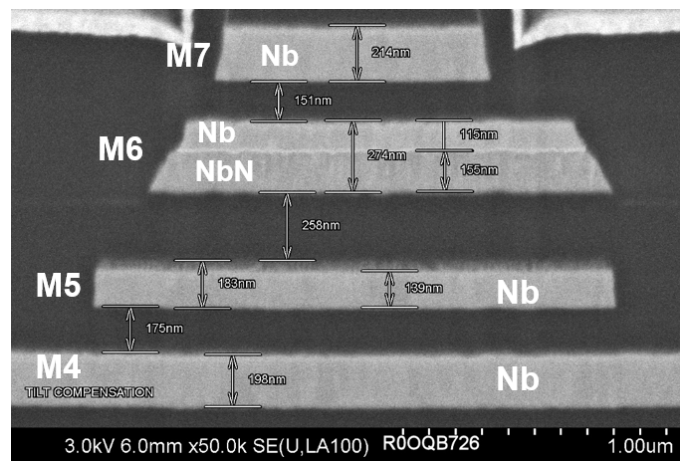

Fig. 14. Cross section of the upper four layers, M4 to M7, of the full 9-superconductor layer SFQ5ee process with a NbN(150)/Nb(100) bilayer deposited as the layer M6; the actual thicknesses of the layers are shown.

etching the top Nb of the bilayer in the regions where NbN inductors are going to be placed. High etch selectivity was achieved due to the use of an etch-stop layer and an optical emission spectrometer for the etch end-point detection. After selectively etching the top Nb in the unmasked areas, the second photolithography is done using a clear field M6 and positive PR. Then, the bilayer is etched off from all unmasked area in one etching step, stopping on the underlaying $SiO_2$. This two-step patterning forms variable thickness composite inductors consisting of thin NbN and thick NbN/Nb parts and/or uniform-thickness inductors consisting of the NbN/Nb bilayer. After that, the process layer stack was completed similar to the standard SFQ5ee process. A cross section of the top part, layers M4 to M7, of the processed layer stack, showing a patterned NbN/Nb bilayer on the level M6, is given in Fig. 14.

## IV. ELECTRICAL TEST RESULTS

### A. Penetration Depths Measurements and Inductance of NbN Microstrips and Striplines with Nb Ground Planes

Magnetic field penetration depth in NbN films was measured using a dielectric (sapphire) resonator technique, unpatterned films, and a setup described in [21]. Inductance and mutual inductance of NbN inductors in multilayered circuits were measured using a SQUID-based method [31] described in detail in [18], and an integrated circuit developed in [17, Fig. 4]. The penetration depth, $\lambda_{NbN}$ in NbN films deposited at different condition by the reactive sputtering and by the PECVD from the metal-organic precursor are given in Fig. 14.

In microscopic theory of superconductivity, the penetration depth in weak coupling superconductors with short mean free path of electrons is given by [32], [33]

$$\lambda(T) = \left(\frac{\hbar\rho}{\mu_0\pi\Delta\tanh\frac{\Delta}{2k_BT}}\right)^{1/2},$$ (3a)

$$\lambda_{BCS} = 0.1048(\rho/\eta_\Delta T_c)^{1/2} \text{ at } T \ll T_c,$$ (3b)

where $\Delta$ is the superconducting energy gap, $\rho$ the normal state resistivity, $\hbar$ the reduced Planck's constant. In (3b), $\lambda_{BCS}$ is in micrometers if $\rho$ is in $\mu\Omega\cdot$cm; $\eta_\Delta = \Delta/1.764k_BT_c$ is the $\Delta/k_BT_c$ ratio of the superconductor to the BCS theory ratio of 1.764 [34]. Dependence (3b) at $\eta_\Delta = 1$ is shown in Fig. 14 by a solid line.

Penetration depth values from ~100 nm in epitaxial single crystal NbN films up to ~600 nm in polycrystalline films have been reported [36]–[42], depending on the resistivity, deposition method, substrate material, deposition temperature, and microstructure of the films. These data are also included in Fig. 14.

NbN is considered a strong coupling superconductor with $\Delta/k_BT_c = 2.155$ [35], i.e., $\eta_\Delta = 1.222$. Dependence (3b) with $\eta_\Delta = 1.222$ is also shown in Fig. 14 and corresponds to the lower boundary of the values of the penetration depth expected in NbN films with short mean-free path.

Penetration depth values in the NbN films deposited by reactive sputtering and PECVD in this research, Fig. 14, are noticeably larger than (3b), corresponding to about 25% lower superfluid density than that expected in conventional superconductors with short mean free path [32]–[34]; the difference increases with decreasing $T_c$ of the films by lowering the deposition temperature.

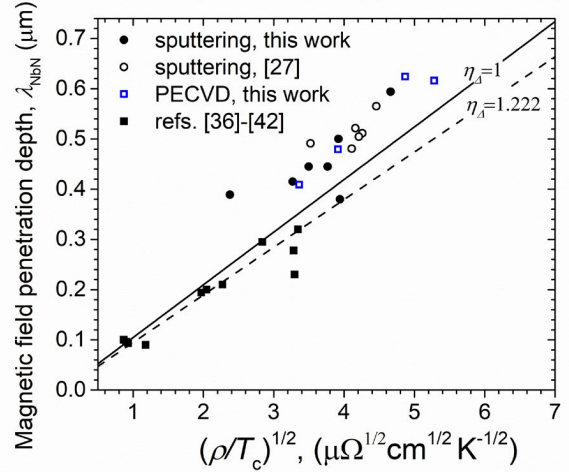

Fig. 14. Magnetic field penetration depth in different NbN films: (●) reactively sputtered at different substrate temperatures and magnetron powers, and measured using the dielectric resonator method; (□) PECVD using the TBTDEN precursor and deposition temperatures from 200 °C to 350 °C, measured using the sapphire resonator method; (○) reactively sputtered, layer M6 in the full 9-metal layer fabrication process, measured using microstrips and stripline inductors with Nb ground plane(s) [27]; (■) deposited by various PVD methods, data from [36]–[42]. Solid and dashed lines – BSC theory dependence (3b) for weakly coupled superconductors with $\Delta/k_BT_c = 1.764$ ($\eta_\Delta = 1$) and $\Delta/k_BT_c = 2.155$ measured on NbN [35] ($\eta_\Delta = 1.222$), respectively.

Results of the inductance measurements on the films integrated into the full nine-superconductor-layer process were presented in [27]. For 200-nm NbN films deposited at 200 °C by reactive sputtering, the extracted penetration depth is $\lambda_{NbN} = 491$ nm. Inductance in all studied microstrip and stripline configurations with Nb ground planes was dominated by the kinetic inductance of the NbN strip. The sheet kinetic inductance $L_K = \mu_0\lambda_{NbN}^2/t$ of 200-nm NbN films was found to be 1.51 pH/sq. Consequently, 100-nm and 50-nm-thick NbN films should have 2× and 4× higher sheet kinetic inductance, about 3.0 pH/sq and 6.0 pH/sq, respectively. These data were used for Table I.

### B. Critical Current of NbN Wires and Comparison to Nb

Another import parameter of superconducting films for implementing them in integrated circuits is their critical current. NbN films deposited by reactive sputtering and incorporated in the full process were patterned using a 248-nm photolithography and HDP etching into long, about 5 mm, "snakes" with the wire widths from 250 nm to 1 μm, shown in Fig. 13. Their critical currents are shown in Fig. 16. For comparison, we give the data obtained on Nb films patterned by the same 248-nm lithography, using a Canon EX4 stepper, and by the 193-nm photolithography, using an ASML scanner.

Due to a much smaller penetration depth, with increasing the linewidth, Nb films transition from the regime $w, t \sim 2\lambda$ where $I_c \propto w$ to the $w \gg \lambda^2/t$ regime where $I_c \propto w^{\frac{1}{2}}$ because of the current crowding near the film edges. NbN films show $I_c \propto w$ dependence in the entire range of linewidths studied because their linewidth is always in the regime $w < 2\,\lambda^2/t$ where the current distribution is uniform. However, the slope of the linear dependence, the critical current density, $J_c$, in NbN films is about nine times lower than that in Nb films.

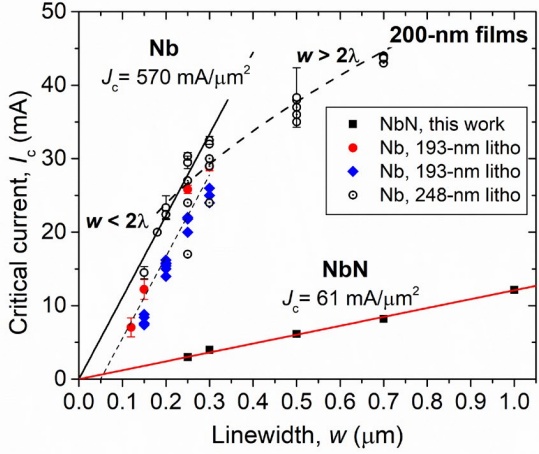

Fig. 16. Critical current of NbN and Nb wires as a function of their linewidth. The films were patterned in meanders with equal linewidth and spacing; the typical length is about 5 mm. NbN films were deposited by reactive magnetron sputtering, Nb films by the magnetron sputtering.

The $J_c$ observed in our Nb films is close to the one expected from the depairing mechanism giving a well-known Ginzburg-Landau (GL) critical current density $J_c^{GL}(4.2\ \text{K}) = \frac{\Phi_0}{3\sqrt{3}\mu_0\xi_{GL}\lambda^2}$. For the Nb films at 4.2 K, $J_c^{GL}(4.2\ \text{K}) \approx 0.59$ A/μm²; see [22] for details. From the GL expression

$$J_c^{GL} = \frac{\Phi_0}{3\sqrt{3}\xi_{GL}L_K t}, \tag{4a}$$

and

$$I_c^{GL} = \frac{0.398w\ [mA\cdot pH]}{\xi_{GL}L_K\ [pH]}, \tag{4b}$$

i.e., the critical current is inversely proportional to the film's sheet kinetic inductance, where $\xi_{GL}$ is Ginzburg-Landau coherence length, in the same units as $w$, and kinetic inductance is in pH. The measured kinetic inductance of the 200-nm-thick NbN films $L_K$ =1.51 pH/sq [27] is 30 times higher than the $L_K$ of Nb films with the same thickness. Therefore, the depairing critical current of NbN films should be much smaller than that of Nb films simply because the superfluid density in NbN films is lower than that in Nb films.

### C. Critical Current of Vias Between NbN and Nb Layers and Between Two NbN Layers

In the full process with NbN or NbN/Nb layer M6, critical currents of the two types of vias are important: vias I5 between the M6 and the bottom Nb layer M5, the base electrode of the JJs; and via I6 between the upper layer M7 and the M6 layer. In all cases, the square vias are etched in the SiO₂ dielectric and filled in by the upper metal layer during its deposition. We varied the thickness of the individual layer in the M6 bilayer, $t_{NbN}$ and $t_{Nb}$, and the material of the M7 layer to investigate critical currents of the vias. The layer M7 was deposited as either a 200-Nb layer or a NbN layer. All NbN layers were deposited at 200 °C by reactive sputtering, as described in III.E. PECVD deposition of NbN layer has not been implemented in the full process layer stack.

Critical currents of multiple chains of 1848 vias between the NbN/Nb and Nb layers, and between NbN/Nb and NbN layers were measured across multiple wafers and the average values are given in Table III. For comparison, $I_c$ of Nb-to-Nb vias, $t_{NbN}$ =0, of the same types are also given.

The critical current of all I5 vias when a thin NbN layer makes contact to the Nb bottom layer, the critical current of NbN-to-Nb via is reduced with respect to Nb-Nb visa by about a factor of ten, similarly to the reduction of the critical current of NbN wire in comparison to Nb wires shown in Fig. 16. This means that the critical current of the NbN layer on the via sidewalls, which depends on the $t_{NbN}$ on the sidewalls, mainly determines the via critical current. The sidewall coverage depends on the via depth and slope. In the process, the via slope is about 75°, giving about 50% reduction in the sidewall film thickness. Then, the via critical current can be estimated as $I_{c,via} \approx 0.5t_{NbN}w_{via}J_c$ giving $I_{c,via} \approx 3.2$ mA for $w_{via} = 0.7$ μm, $t_{NbN}$ =150 nm, and $J_c$ =61 mA/μm² given in Fig. 16, in general agreement with the data in Table III. On the other hand, when the contact is made by the top Nb layer to the NbN or NbN/Nb bilayer, as in the I6 vias, the $I_{c,via}$ is about a factor 2× higher than for the I5 vias because there is no NbN thickness reduction in these cases.

| Via type and size, $w_{via}$ | Via depth (nm) | M6 NbN/Nb thicknesses (nm) | M7 NbN/Nb thicknesses (nm) | $I_{c,via}$ (mA) |
|---|---|---|---|---|
| I5 0.5 μm[a] | 280 | 0/200 | 0/200 | 25 |
| I5 0.5 μm | 280 | 150/100 | 0/200 | 2.5 |
| I5 0.5 μm | 280 | 130/70 | 0/200 | 2.2 |
| I5 0.5 μm | 280 | 100/100 | 250/0 | 2.3 |
| I5 0.7 μm | 280 | 0/200 | 0/200 | 35 |
| I5 0.7 μm | 280 | 200/0 | 0/200 | 2.8 |
| I5 0.7 μm | 280 | 150/100 | 0/200 | 5.6 |
| I5 0.7 μm | 280 | 130/70 | 0/200 | 4.4 |
| I5 0.7 μm | 280 | 100/100 | 250/0 | 5.9 |
| I6 0.5 μm[b] | 200 | 0/200 | 0/200 | 49 |
| I6 0.5 μm | 200 | 150/100 | 0/200 | 6.0 |
| I6 0.5 μm | 200 | 130/70 | 0/200 | 3.1 |
| I6 0.5 μm | 200 | 100/100 | 250/0 | 6.6 |
| I6 0.7 μm | 200 | 0/200 | 0/200 | >50 |
| I6 0.7 μm | 200 | 150/100 | 0/200 | 35.1 |
| I6 0.7 μm | 200 | 200/0 | 0/200 | 12.8 |
| I6 0.7 μm | 200 | 130/70 | 0/200 | 26.8 |
| I6 0.7 μm | 200 | 100/100 | 250/0 | 8.0 |

[a] Via I5 is between a 135-nm thick Nb base electrode of Josephson junctions, layer M5, and layer M6 – a NbN/Nb bilayer with different thicknesses $t_{NbN}$ and $t_{Nb}$, or a 200-nm layer for comparison. The bottom NbN layer of the M6 bilayer contacts Nb of the M5 layer. All data are for 1848-via chains.

[b] Via I6 is between layers M6 and M7. The latter is either a Nb layer or NbN layer for comparison. The M7 metal contacts the top layer of the M6 bilayer.

TABLE IV
CRITICAL CURRENTS OF HIGH-TEMPERATURE ANNEALED NbN WIRES AND
VIAS BETWEEN TWO NbN LAYERS

| Parameter | SiO$_2$ deposited at 150 °C | SiO$_2$ deposited at 400 °C |
|---|---|---|
| $T_c$ (K) | 15.9 | 15.5 |
| $I_c$ of 0.7-µm wires (mA) | 5.2 | 4.2 |
| $I_c$ of 1848 0.7-µm I6 vias between 200-nm NbN layer M6 and 200-nm NbN layer M7 (mA) | 7.0 | 6.1 |

The measured critical currents of 0.5- and 0.7-µm vias involving NbN layers, although significantly lower than of Nb vias, appear to be sufficient for the use in integrated circuits. However, the minimum size and $I_{c,via}$ of the vias will ultimately determine the achievable scale of integration; see Fig. 2. Therefore, it remains to be investigated if $I_{c,via}$ can be increased and the via size reduced using the PECVD-filled vias or a stud-via, pillar-via, process developed for Nb vias in [49] and implemented in the full 9-Nb-layer process in [20].

### D. Temperature Stability

Another important issue is temperature stability of NbN wires and vias at elevated temperatures in atmosphere of gases used for PECVD of SiO$_2$. We investigated superconducting properties of two NbN layers, M6 and M7, and I6 vias between them, etched in a 200-nm SiO$_2$ dielectric. The temperature stability was investigated using SiO$_2$ deposition, which lasts about 30 min, at high temperatures over the patterned NbN wires and formed vias. The obtained results are given in Table IV. Only a minor degradation of NbN films has been detected at high dielectric deposition temperatures of 400 °C. For comparison, similar Nb structures would not be superconducting after this treatment.

### V. CONCLUSION

The goals of this research were to show our progress towards implementation of NbN kinetic inductors in a multilayered fabrication process node titled SFQ7ee and containing ten fully planarized superconducting layers.

Other convenient material for incorporation in our process and integration with Nb/Al-AlO$_x$/Nb junctions could be NbTiN films [40]–[46], which have a smaller penetration depth than NbN, in the range from 230 nm to 360 nm, depending on the deposition conditions.

Full implementation of NbN/Nb or NbTiN/Nb bilayer inductors should provide for a 10× increase in the inductor number density and hence significantly increase integration scale of superconductor electronics. We hope that, by using one or two layers of kinetic inductors and one layer of self-shunted Josephson junctions, we can increase the integration scale close to a 100 million JJs per cm$^2$. The other means of increasing the integration scale and circuit density further would be by increasing the number of Josephson junction layers to two and beyond.

Some progress in this direction has already been demonstrated [47], [48].

### ACKNOWLEDGMENT

We are grateful to Vasili Semenov for numerous discussions of bilayer inductors.

This research was based upon work supported by the Under Secretary of Defense for Research and Engineering via Air Force Contract No. FA8702-15-D-0001. Any opinions, findings, conclusions or recommendations expressed in this material are those of the authors and do not necessarily reflect the views of the Under Secretary of Defense for Research and Engineering and should not be interpreted as necessarily representing the official policies or endorsements, either expressed or implied, of the U.S. Government. Delivered to the U.S. Government with Unlimited Rights, as defined in DFARS Part 252.227-7013 or 7014 (Feb 2014). Notwithstanding any copyright notice, U.S. Government rights in this work are defined by DFARS 252.227-7013 or DFARS 252.227-7014 as detailed above. Use of this work other than as specifically authorized by the U.S. Government may violate any copyrights that exist in this work. The U.S. Government is authorized to reproduce and distribute reprints for Governmental purposes notwithstanding any copyright annotation thereon.